\documentclass[pre,floatfix,superscriptaddress,nofootinbib, longbibliography,twocolumn,final]{revtex4-1}

\usepackage{graphicx}
\usepackage[utf8]{inputenc}
\usepackage[english]{babel}
\usepackage{microtype}

\usepackage[boxruled]{algorithm2e}
\SetAlFnt{\footnotesize}
\SetAlCapFnt{\small}
\SetAlCapNameFnt{\small}

\usepackage{amsmath}
\usepackage{amsfonts}
\usepackage{amsthm}
\usepackage{amssymb}
\usepackage{bm} % write every math-letter boldface using \bm, including greek characters..

\usepackage{tabularx}
\usepackage{booktabs} % for much better looking tables
\usepackage{array} % for better arrays (eg matrices) in maths
\usepackage{dsfont}
\usepackage[normalem]{ulem}

\usepackage[colorlinks, breaklinks,hyperfootnotes = false,citecolor = red]{hyperref} %colorlinks
\usepackage[usenames]{color}

\usepackage[capitalise]{cleveref} % for clever refs

\newcommand{\dav}[0]{\ensuremath{\langle \delta \rangle}}

\newcommand{\SNRf}[0]{\ensuremath{\textrm{SNR}_\text{fine}}}
\newcommand{\SNRc}[0]{\ensuremath{\textrm{SNR}_\text{coarse}}}
\newcommand{\SNRl}[0]{\ensuremath{\textrm{SNR}_\lambda}}
\newcommand{\SNRcond}[0]{\ensuremath{\textrm{SNR}_\text{nested}}}

\begin{document}
\title{Detectability of hierarchical communities in networks}

\author{Leto Peel}
\email{l.peel@maastrichtuniversity.nl}
\affiliation{Department of Data Analytics and Digitalisation, School of Business and Economics, Maastricht University}
\author{Michael T. Schaub}
\email{schaub@cs.rwth-aachen.de}
\affiliation{Department of Computer Science, RWTH Aachen University}

\begin{abstract}
    We study the problem of recovering a planted hierarchy of partitions in a network.
    The detectability of a single planted partition has previously been analysed in detail and a phase transition has been identified below which the partition cannot be detected. 
    Here we show that, in the hierarchical setting, there exist additional phases in which the presence of multiple consistent partitions can either help or hinder detection. 
    Accordingly, the detectability limit for non-hierarchical partitions typically provides insufficient information about the detectability of the complete hierarchical structure, as we highlight with several constructive examples. 
\end{abstract}

\maketitle

Community detection has been one of the core topics in the study of complex networks for the last two decades~\cite{Fortunato2010}. 
The foundational work of Girvan and Newman~\cite{girvan2002community} identified the presence of locally dense substructures within the globally sparse structure of many real-world networks. 
Furthermore, they found that nodes within these dense substructures, or \textit{communities}, tended to share certain properties.  
The continued interest in detecting communities in networks lies in the fact that decomposing a network into groups of similarly connected nodes often provides an insight into structure and function of a complex system~\cite{Fortunato2010}.

Following the example of Newman and Girvan~\cite{Newman2004} most community detection algorithms aim at identifying groups of nodes that are more densely connected to each other than to the rest of the network.
Different notions of community structure have since been introduced~\cite{Schaub2017} to address specific network systems, but still involve the core idea of a partition of a network into groups of nodes.
Here we focus on a particular conceptualisation of community structure based on the so-called stochastic block model~\cite{holland1983stochastic}. 
The stochastic block model posits that nodes within the same community share the same set of probabilities to connect to nodes in other groups.
This formulation of community structure in terms of a probabilistic generative model enables the precise theoretical study of community detection from a computational and statistical perspective.
Many important theoretical advances have been made concerning community detection in networks based on this model~\cite{abbe2018community}.

In their seminal work~\cite{decelle2011asymptotic}, Decelle et al.\ conjectured that a theoretical detectability limit exists, below which no efficient (i.e., polynomial time complexity) community detection algorithm can perform better than a random assignment. 
This limit has been formulated in terms of the planted partition model, a random graph model in which the nodes are split into $k$ equal-sized communities.
Nodes in the same community are then connected with probability $p_{\rm in}$ and nodes in different communities are connected with probability $p_{\rm out}$.
The (correctly ordered) adjacency matrix of the graph can thus be described by $k^2$ blocks, where the diagonal blocks can be described by $\delta_{\rm in}= np_{\rm in}/k$, the expected number of edges a node shares within the same community, and the off-diagonal blocks  can be described by $\delta_{\rm out}= np_{\rm out}/k$, the expected number of edges between a node and other communities  
(for simplicity we allow for self-loops).
For a network with $n$ nodes and $k$ groups of equal size, the condition for detectability can then be written as:
\begin{equation}\label{eq:KS_threshold}
    \left (\delta_{\rm in} - \delta_{\rm out} \right )^2 \ge \dav \enspace ,
\end{equation}
where $\dav =\delta_{\rm in} + (k-1)\delta_{\rm out}$ is the expected degree of any node in the graph.
The detectability limit is reached when Eq.~\eqref{eq:KS_threshold} holds with equality.

The conjecture that there is such a detectability limit was first proven by a series of papers for the case $k=2$: Masoulie et al.~\cite{massoulie2014community} and Mossel et al.~\cite{Mossel2018} independently provided sharp upper bounds for the detectability threshold.
Mossel et al.~\cite{mossel2015reconstruction} then provided a matching lower bound, thereby confirming the conjectured threshold in Eq.~\eqref{eq:KS_threshold} holds for $k=2$.
Since then we  have seen numerous theoretical advances extending these results. 
Surveys of these advances and a historical overview are presented by Abbe~\cite{abbe2018community} and Moore~\cite{moore2017computer}.

Until now, the focus of all these theoretical developments has been on the detection of a single partition of a network. 
Here we consider a setting where \emph{multiple} consistent partitions exist at different resolutions, and we are interested in the detectability of \textit{all} of these partitions.
More precisely, we are interested in networks that contain hierarchical groupings, in which communities are further divided into subgroups. 
Using theoretical arguments and computational experiments we will show that, compared to detecting a single partition, an even richer picture emerges when considering the detectability of these hierarchical groupings.

\section{Detectability of a network partition}
We start by briefly reviewing the situation for recovering a single partition from a graph generated from a stochastic block model (SBM).
For simplicity, we will focus on undirected networks. 
The SBM is a probabilistic generative model from which we can sample networks of a finite size with a specified community structure. 
The SBM assigns each of the $n$ nodes in a network to one of $k$ groups of nodes. 
Here we assign nodes to groups according to a uniform distribution such that each node is assigned to any group with probability $1/k$. 
Each link $A_{ij}$ of the adjacency matrix $\bm A \in \{0,1\}^{n\times n}$ is (up to the symmetry of the matrix) an independent Bernoulli random variable with a parameter that depends only the group assignments of nodes $i$ and $j$. 
To compactly describe the model, we collect all these Bernoulli parameters in a symmetric affinity matrix $\bm \Omega \in [0,1]^{k \times k}$, in which the element $\Omega_{rs}$ represents the probability of an edge occurring between a node in group $r$ and a node in group $s$. 
We represent the group assignment using a partition indicator matrix $\bm H \in \{0,1\}^{n\times k}$ with entries $H_{ij} = 1$ if node $i$ belongs to group $j$ and $H_{ij}=0$ otherwise.
The expected adjacency matrix under the SBM is written as
\begin{equation}\label{E:expected_G}
  \mathbb{E}[\bm A | \bm H, \bm \Omega] = \bm H \bm \Omega \bm H^\top \enspace .
\end{equation}

A special case of the SBM is the planted partition model, in which the affinity matrix comprises two parameters: $\Omega_{rr} = p_\text{in}$, representing the within group connectivity, and $\Omega_{rs} = p_\text{out}$, for $r\neq s$, representing the across group connectivity.
We can write the affinity matrix for a graph with $n$ nodes and a $k$-group planted partition as:
\begin{equation}\label{eq:ppmdef}
    \bm \Omega = p_\text{in}\bm I_k  + p_\text{out}(\bm 1_k^{} \bm 1_k^\top - \bm I_k) \enspace ,
\end{equation}
where $\bm I_k$ and $\bm 1_k$ are the $k$-dimensional identity matrix and the $k$-dimensional vector of all-ones, respectively.

The detectability limit as originally expressed in terms of Eq.~\eqref{eq:KS_threshold}~\cite{decelle2011asymptotic,moore2017computer,abbe2018community} is based on a planted partition model with equally sized groups in the limit where $n\rightarrow \infty$ and the connection probabilities $p_\text{in}$ and $p_\text{out}$ scale as $1/n$. 
Consequently, each node has a constant expected degree and the generated networks are sparse. 
Rearranging Eq.~\eqref{eq:KS_threshold}, we can write an expression for the signal-to-noise ratio for a planted partition (SNR$_{\rm{PP}}$),
\begin{equation}
  \textrm{SNR}_{\rm{PP}} = \frac{\left( \delta_{\rm in} - \delta_{\rm out}\right)^2}{\dav} \enspace .
  \label{eq:SNRPP}
\end{equation}
A lower SNR$_{\rm{PP}}$ indicates that the partition is harder to detect. Below the limit SNR$_{\rm{PP}} = 1$, the partition becomes undetectable.

For the more general case of an SBM (with affinity matrix $\bm \Omega$ and group indicator matrix $\bm H$), Abbe and Sandon~\cite{abbe2015detection} defined a more general signal-to-noise ratio as a function of the eigenvalues of the matrix  $\bm Q = \bm \Omega \bm H^\top \bm H$. 
With the eigenvalues $|\lambda_1| \geq |\lambda_2| \geq \ldots \geq |\lambda_k|$ ordered in non-increasing magnitude, the signal-to-noise ratio ($\SNRl$) is given as:
\begin{equation}\label{eq:SNR_lambda}
    \SNRl = \frac{\lambda_2^2}{\lambda_1} \enspace .
\end{equation}
Here $\SNRl = 1$ corresponds to the detectability limit and is equivalent to \cref{eq:SNRPP} for the planted partition model. 
A higher $\SNRl$ implies that it is easier to recover the planted partition.
However, in the following section we will see that, for a hierarchical community partition, $\SNRl$ is not really informative about the detectability of the complete hierarchy.

We remark that as $\bm Q$ is a non-negative matrix, $\lambda_1$ will always be non-negative according to Perron--Froebenius theory and $\SNRl$ is therefore always non-negative.
Moreover, it can be shown that the eigenvalues of $\bm Q$ correspond precisely to the nonzero eigenvalues of $\mathbb{E}[\bm A | \bm H, \bm \Omega]$.

\section{Detectability of hierarchical partitions}
We now consider a setup in which the network under consideration contains a hierarchical planted partition: first a two-level hierarchy and then a three-level hierarchy.

\subsection{Two-level hierarchies}
We begin with a hierarchy that consists of a coarse partition and a fine partition.
The coarse partition contains $k_1$ equally-sized groups. 
Each group in the coarse partition is further partitioned into $k_2$ equally-sized groups to create the fine partition that contains $k=k_1 k_2$ groups in total. 
We will focus on the case that $k_1=k_2=2$ and a three-parameter affinity matrix:
\begin{equation}\label{eq:hier_SBM_affinity}
\bm \Omega = \frac{1}{n}
\begin{bmatrix}
  a & b & c & c \\
  b & a & c & c \\
  c & c & a & b \\
  c & c & b & a 
\end{bmatrix} \enspace ,
\end{equation}
where $a,b$ and $c$ are parameters that do not depend on $n$.
More generally, however, we can express the rescaled matrix $\bm Q = n\bm \Omega/k$ for a two-level hierarchy as:
\begin{subequations}\label{eq:hier_SBM_affinity_general}
\begin{align}
    \bm Q = \frac{1}{k} \left [\bm I_{k_1} \otimes \bm \omega  + c (\bm 1_{k_1}^{} \bm 1_{k_1}^\top -\bm I_{k_1}) \otimes \bm 1_{k_2}^{} \bm 1_{k_2}^\top \right ] \enspace ,\\
    \text{where}\quad \bm \omega = a \bm I_{k_2}  + b (\bm 1_{k_2}^{}\bm 1_{k_2}^\top - \bm I_{k_2}) \enspace .
\end{align}
\end{subequations}
The corresponding eigenvalues of $\bm Q$ take three distinct values $\alpha_1, \alpha_2$ and $\alpha_3$, with different multiplicities:
\begin{align*}
  \alpha_1 & = [a + (k_2-1)b + (k_1-1)k_2 c]/k & (\text{once}) & \\
  \alpha_2 & = [a + (k_2-1)b - k_2 c]/k  & (k_1-1~\text{times}) & \\
  \alpha_3 & = (a -b)/k & (k-k_1~\text{times}) & \enspace . 
\end{align*}
 The eigenvalue with the largest magnitude will always be $\lambda_1 = \alpha_1$. 
 The second largest eigenvalue $\lambda_2$ will depend on the parameters $b$ and $c$.
 Since $\alpha_2 - \alpha_3 = (b-c)/k_1$, the second largest eigenvalue will be $\alpha_2 >\alpha_3$ when $b>c$ and $\alpha_3 >\alpha_2$ when $c>b$.  
 In the degenerate case where $b=c$, both eigenvalues will be equal and the hierarchical community structure simply reduces to a planted partition into $k$ groups. 

To see why $\SNRl$ cannot be informative about the complete hierarchy, we can inspect the subspaces associated with these eigenvalues.  
The $(k_1-1)$ dimensional subspace $\mathcal A_2 \subset \mathbb{R}^{k}$ associated with the eigenvalue $\alpha_2$ can be spanned by eigenvectors of the form:
\begin{equation}
  v^{(s)}_t = \left\{
  \begin{aligned}
	1 \quad & \textrm{if} \quad (s-1)k_2 & <  t & \leq & sk_2 \\
	-1 \quad  & \textrm{if} \quad sk_2 & <  t & \leq & (s+1)k_2 \\
	0 \quad & \textrm{otherwise} \\
  \end{aligned}
  \right.
  \enspace ,
\end{equation} 
where $v^{(s)}_t$ is the $t$-th element of the $s$-th eigenvector $\bm v^{(s)}$ with eigenvalue $\alpha_2$, where the index $s$ ranges from $1 \leq s \leq (k_1-1)$.
These eigenvectors relate to differences in link density between blocks at the coarse level, i.e., each entry of the vector $\bm Q \bm v^{(s)}$ takes the expected sum of edges in one coarse block and subtracts the sum of edges in another coarse block. 
Similarly the $(k -k_2)$ dimensional subspace $\mathcal A_3 \subset \mathbb{R}^{k}$ associated with the eigenvalue $\alpha_3$ can be spanned by shifts of $[1,-1,0,\ldots,0]^\top$ that analogously relate to the  differences in link density between the fine-grained blocks.
Hence, $\SNRl$ will be determined entirely by the differences at \textit{either} the coarse \textit{or} the fine hierarchical level, with no consideration to the nested structure of the problem.
In fact, in the prototypical hierarchical case in which $b\gg c$, the value of $\textrm{SNR}_\lambda$ will be driven entirely by the coarse group structure and the subgroup structure will be irrelevant, which is clearly not ideal.

To analyze the detectability of the partitions at each level of the hierarchy, we introduce two formulae for the signal-to-noise ratio that are commensurate with~\cref{eq:SNRPP}. 
First, we consider the signal-to-noise ratio $\SNRc$ to capture the detectability of the coarse partition at the first level of the hierarchy,
\begin{equation}
    \SNRc = \dfrac{\left(\delta^{(1)}_{\textrm{in}} - \delta^{(1)}_{\textrm{out}}\right)^{2}}{\dav} \enspace ,
    \label{eq:snrc}
\end{equation}
where the degree parameters $\delta_\text{in}^{(1)}$ and $\delta_\text{out}^{(1)}$  are given by:
\begin{align}
    \delta^{(1)}_{\textrm{in}} & = a/k + (k_2-1)b/k \enspace , \\
    \delta^{(1)}_{\textrm{out}} & = k_2 c/k = c/k_1 \enspace ,
\end{align}
and the average degree of the network is calculated as $\dav = [a + (k_2-1)b + (k_1-1)k_2 c]/k =\alpha_1$, which corresponds to $\dav = \delta^{(1)}_{\textrm{in}} + (k_1-1) \delta^{(1)}_{\textrm{out}}$ similar to the planted partition case.

Second, we define the signal-to-noise ratio $\SNRf$ to capture the detectability of the fine partition at the second level of the hierarchy,
\begin{equation}
    \SNRf = \dfrac{\left(\delta^{(2)}_{\textrm{in}} - \delta^{(2)}_{\textrm{out}}\right)^{2}}{\dav} \enspace ,
    \label{eq:snrf}
\end{equation}
where the corresponding in- and out-degree parameters are given by:
\begin{align}
    \delta^{(2)}_{\textrm{in}} & = a/k \\
    \delta^{(2)}_{\textrm{out}} & = \frac{1}{k-1} [(k_2-1)b/k+(k-k_2)c/k] \enspace .
\end{align}

\begin{figure}
  \includegraphics[width=\columnwidth]{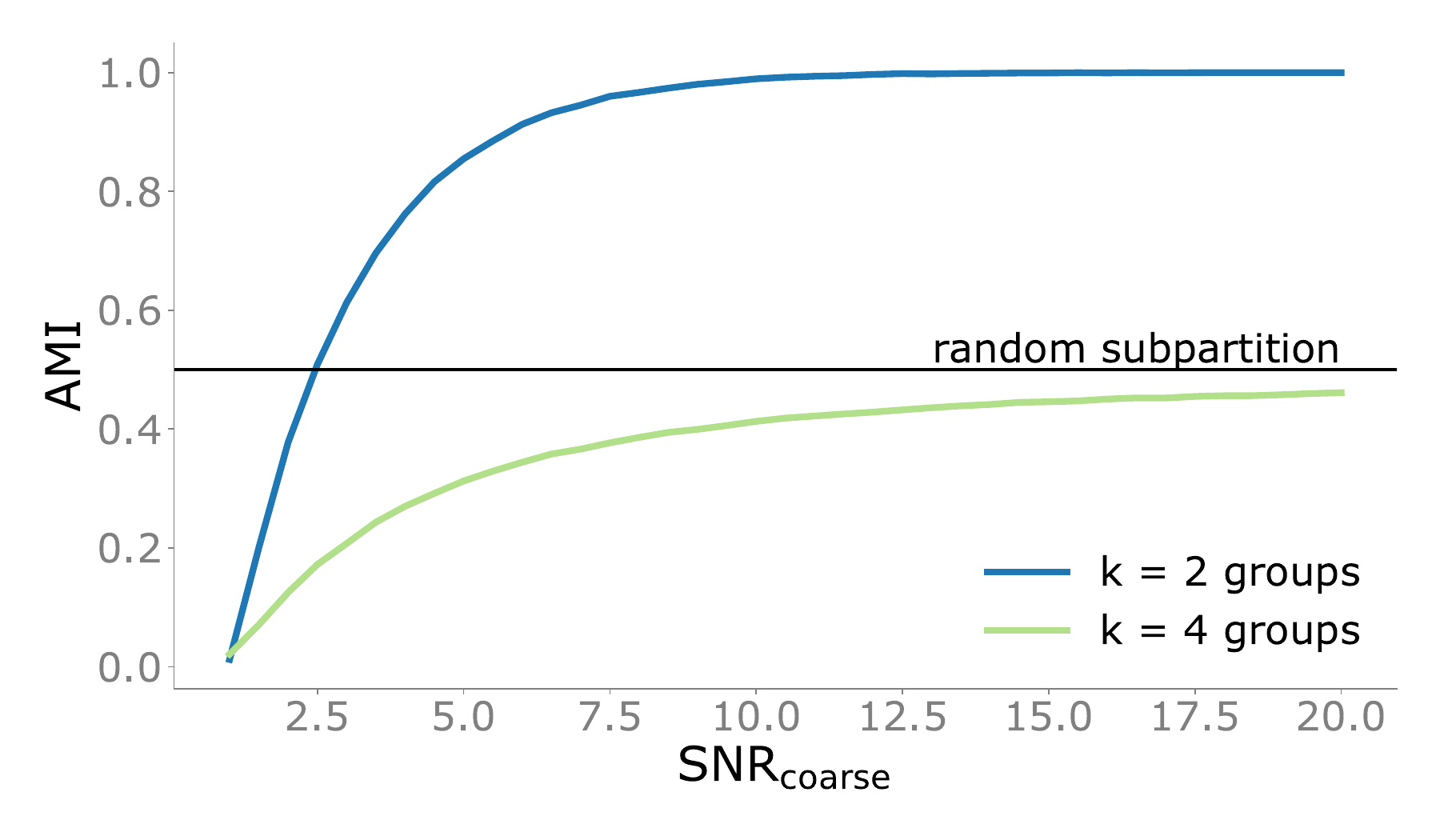}
  \caption{Increasing the SNR of the coarsest partition does not improve the finest partition. Fixing $\SNRf = 1$ and varying $\SNRc$, we measure the performance using adjusted mutual information (AMI). We see that the performance of recovering the partition into 4 groups at second level of the hierarchy tends to the performance of random bipartitions of the groups at the first level.}
  \label{fig:fix_snr}
\end{figure}

Using \cref{eq:snrc,eq:snrf} above, we can set the SNR for the coarse and fine partitions separately, within the constraints of preserving the mean degrees. 
\Cref{fig:fix_snr} illustrates this point and shows the partition recovery performance for a planted hierarchy with $k_1=k_2=2$ and $k=4$ groups. 
Performance is based on spectral clustering using the Bethe Hessian~\cite{Saade2014}, where the number of groups are known. 
(In all experiments we use networks with $n=4^7$ nodes.)
This approach has previously been demonstrated to have near-optimal partition recovery performance right down to the detectability limit. 
Partition recovery is measured by means of the adjusted mutual information ($\text{AMI}$)~\cite{Vinh2010}, for which $\text{AMI}=0$ indicates performance as good as random guessing and $\text{AMI}=1$ equals perfect recovery.
In this experiment the fine partition is fixed at the limit of detectability (fix $\SNRf=1$), while we vary the detectability of the coarse partition (vary $\SNRc$).
We see that as we increase $\SNRc$ the recovery performance increases for both partitions.
However, while we can eventually reach perfect recovery for the coarse partition, the performance on the fine partition is bounded by a performance equivalent to choosing a random subpartitions of the coarse partition. 

This shows numerically that the general signal-to-noise ratio $\SNRl$ does not give us an indication here that the fine split is effectively unrecoverable.
Indeed $\SNRl$ is here equal to the signal-to-noise ratio of the more dominant coarse partition $\SNRc=\SNRl$. 
To analyse the interplay between the two hierarchical partition levels in more details, we parametrize the ratio of the two signal to noise ratios via the ratio $\gamma := \SNRf / \SNRc$.
Considering the regime in which $\delta_\text{in}^{(1)} \ge \delta_\text{out}^{(1)}$ and $\delta_\text{in}^{(2)} \ge \delta_\text{out}^{(2)}$, we can compactly reparametrize the affinity matrix [\cref{eq:hier_SBM_affinity_general}] in terms of $\gamma$, $\SNRc$, and the average degree~$\dav$.
Using this parametrization we can investigate the hierarchical detection problem in more detail. 

\begin{figure}
  \includegraphics[width=\columnwidth]{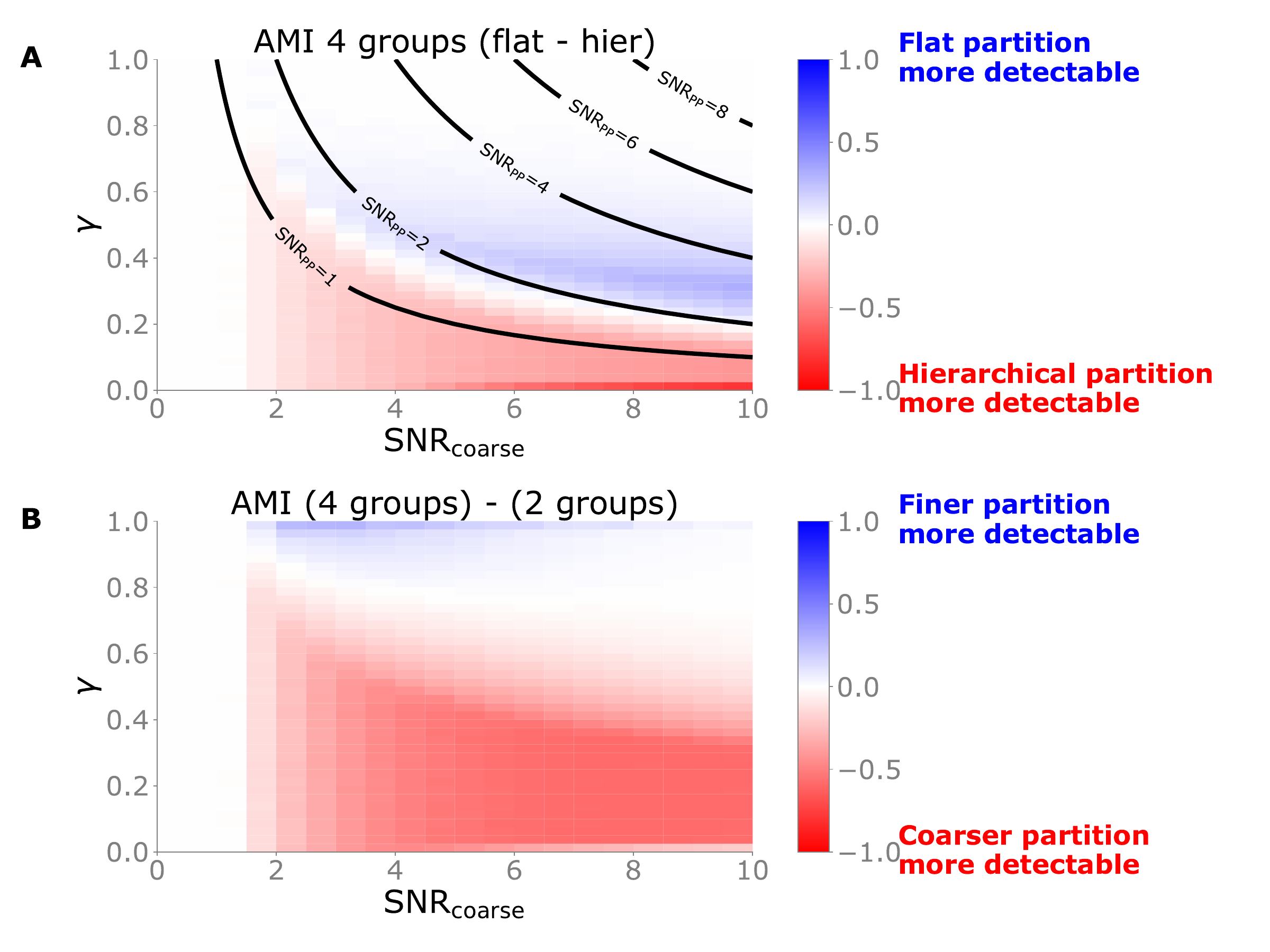}
  \caption{Performance of recovering hierarchical planted partitions. 
  (A) Difference in Adjusted mutual information (AMI) scores for recovering flat partitions versus hierarchical partitions as a function of the coarse signal-to-noise ratio ($\SNRc$) and the ratio $\gamma = \frac{\SNRf}{\SNRc}$. 
  (B) Difference in AMI scores for recovering the finest partition versus the coarsest partition in the hierarchy.
  }\label{fig:2}
\end{figure}

\Cref{fig:2} displays the performance of recovering the fine partition, relative to a non-hierarchical planted partition into four groups (\cref{fig:2}A) and relative to the coarse partition of the hierarchy (\cref{fig:2}B), as we vary $\gamma$ and $\SNRc$. 
Specifically, in \Cref{fig:2}A we compare the recovery of a ``flat'' non-hierarchical planted partitions vs a hierarchical partitions with the same number of groups.
We set SNR$_{\rm{PP}} = \SNRf$ (indicated by contours) to ensure a fair comparison. 
We see that for low values of SNR$_{\rm{PP}}$ the hierarchical partition is easier to recover as the presence of the coarse partition provides partial information about the fine partition.
However, as we increase SNR$_{\rm{PP}}$, we encounter a region where the hierarchical partition becomes harder to detect than the flat planted partition.
Once both $\SNRc$ and $\SNRf$ are sufficiently high (around $\text{SNR}_\text{PP} \geq 6$) we find that the performance of flat and hierarchical partitions becomes equal.

In \Cref{fig:2}B we compare the performance of recovering the fine partition relative to recovery of the coarse partition. 
As $\gamma \leq 1$, it means that $\SNRf \leq \SNRc$ and so it is unsurprising to see that the coarser partition is often easier to detect than the finer partition. 
However, it may be less intuitive to see that as $\gamma \rightarrow 1$ the finer partition becomes more detectable, particularly when $\SNRc$ is small.

\begin{figure}
  \includegraphics[width=\columnwidth]{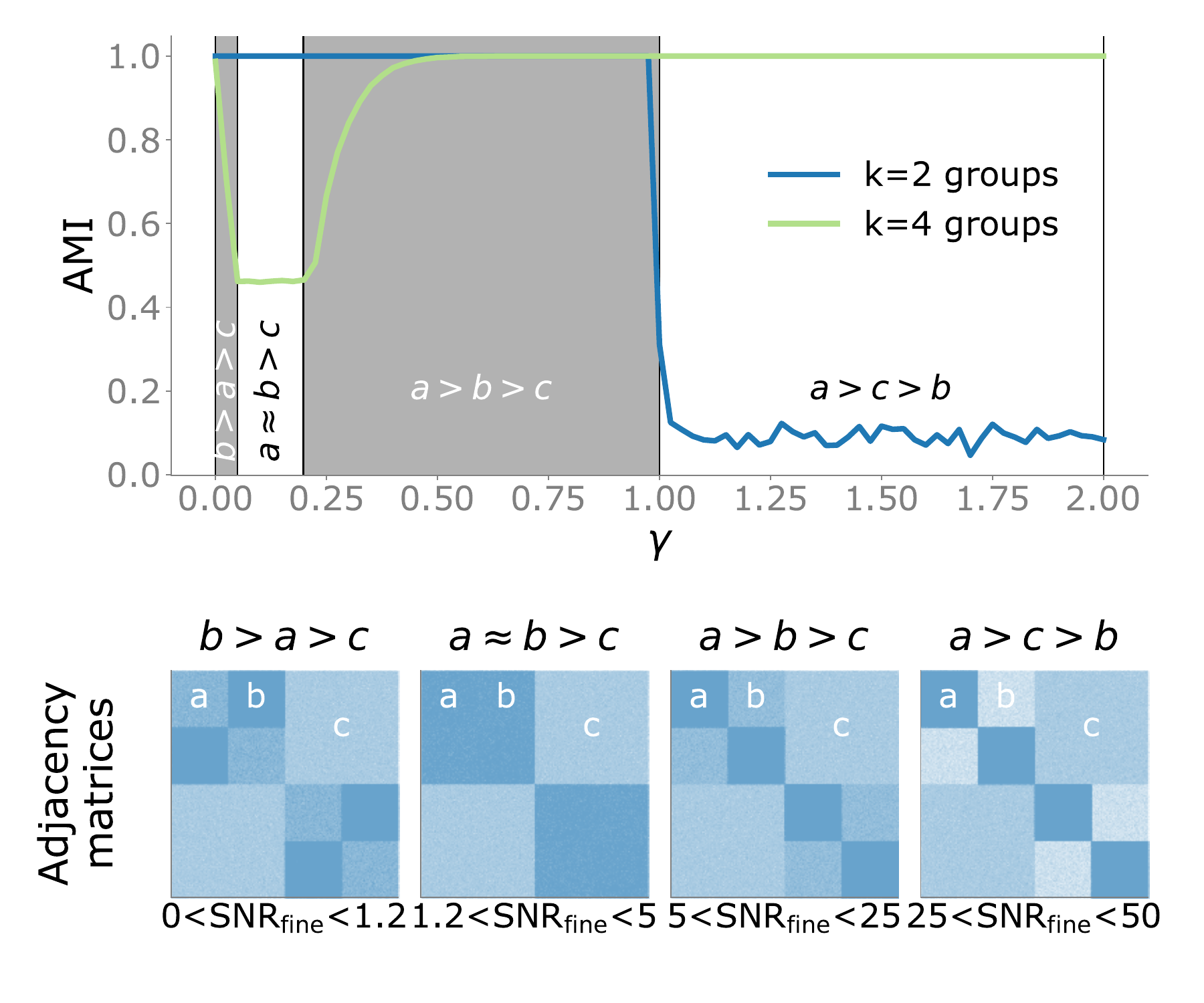}
  \caption{Detection of hierarchical communities as a function of the ratio $\gamma = \frac{\SNRf}{\SNRc}$. We fix $\SNRc = 25$ and as we vary the ratio $\gamma$ the relationship between the three block densities $a, b$ and $c$ transitions across 4 different phases. Note that when $\gamma>1$, spectral clustering prefers the partition $\{1,3\}, \{2,4\}$ (or the equivalent partition $\{1,4\}, \{2,3\}$) instead of $\{1,2\}, \{3,4\}$. At $\gamma=0.2$ ($\rm{SNR}_2 = 5$) there appears to be a detectability threshold for the partition into four groups, below which it is not possible to detect the finest partition any better than randomly partitioning the two groups in the coarser partition into four groups. A spy plot of an example adjacency matrix for each phase is shown along the bottom of the figure.}\label{fig:3}
\end{figure}

To investigate these transitions further, we set up another experiment in which we fixed the coarse signal to noise ratio at $\SNRc=25$, and the average degree to $\dav = 150$ and varied the ratio $\gamma$. 
\Cref{fig:3} shows the results of this experiment and we see that as we vary $\gamma$ we encounter four distinct phases. 

In the first phase ($ 0 < \gamma \lessapprox 0.049$) the parameters of the affinity matrix have the relation $b > a > c$ and both levels of the hierarchy are detectable and the coarse partition is fully recoverable. 
In the second phase ($ {\sim}0.049 < \gamma \lessapprox 0.198$) the parameters $a$ and $b$ are approximately equal, which makes the fine partition undetectable beyond detection of the coarse partition, i.e., we can do no better than randomly partitioning each of the two coarse groups into two subgroups.
We can use this observation to calculate the boundaries of this phase by considering the nested detectability limit of the fine groups within the coarse groups.  
This limit occurs when the nested SNR is equal to 1,
\begin{equation}
  \SNRcond = \frac{(a-b)^2}{k_2(a+b(k_2-1))} = \frac{(a-b)^2}{k^2_2g} = 1 \enspace ,
\end{equation}
where $g = k_1 \delta^{(1)}_{\textrm{in}}$ is average in-degree of the coarse groups. 
Then we can derive the values of the parameters $a, b = g \pm \sqrt{k_1g}$, where the factor $k_1$ adjusts for the fact that the subgraph of one of the coarse groups only contains $n/k_1$ nodes. 
In the third phase (${\sim}0.198 < \gamma < 1$) the parameters of the affinity matrix have the relation $a > b > c$, representing an assortative hierarchy in which the link density is concentrated in the diagonal blocks. 
Both hierarchical partitions are detectable in this phase with high accuracy.
However, as $\gamma \rightarrow 1$ we see a drop in performance in detecting the coarse partition.
This drop in performance occurs because when $\gamma = 1$, there is effectively no coarse partition in the hierarchy, since $b=c=\dav - \sqrt{\dav}$ and so we simply have a planted partition into four groups.
In the final phase ($\gamma > 1$) the parameters of the affinity matrix have the relation $a > c > b$ and both partitions should be detectable because $\SNRc > 1$ and $\SNRf > 1$. 
However, instead we observe poor performance for the spectral clustering in detecting the coarse partition into two groups, even though we can perfectly recover the fine partition.
This observation is a result from a degeneracy in the solution: any three partitions created by pairing up the four groups at the finest level will result in a ``good'' partition at least in terms of spectral clustering. 
Consequently, we have a situation in which communities are detectable, but non-identifiable~\cite{Schaub2020}. In other words any of these configurations could have generated the network. In this situation the community detection algorithm will typically detect the partition with the highest SNR because community detection is usually framed as an optimisation.

When $\gamma<1$, we identify the planted coarse partition as expected, which has $\SNRc^{\gamma<1} =  \frac{(a+b-2c)^2}{a+b+2c}$. However, when $\gamma>1$, we instead find the alternative coarse partition:
\begin{equation}\label{eq:adj_c_greater_b}
\left[\begin{array}{@{}cc|cc@{}}
  a & c & b & c \\
  c & a & c & b \\
  \hline 
  b & c & a & c \\
  c & b & c & a 
\end{array}\right]
 \enspace ,
\end{equation}
which does not correlate with the planted partition. This alternative coarse partition has $\SNRc^{\gamma>1} = \frac{(a-b)^2}{a+b+2c}$. We can easily verify that $\SNRc^{\gamma>1} > \SNRc^{\gamma<1}$ when $c>b$. This result is directly related to the reordering of the eigenvalues of $\bm Q$ because $\alpha_3>\alpha_2$ when $c>b$. Consequently $\alpha_3$ becomes the second largest eigenvalue.

\subsection{More general hierarchies}
We can create hierarchies of greater than two levels by recursively constructing the rescaled matrix $\bm Q = n\bm \Omega/k$. For example, for a three-level hierarchy in which nodes are partitioned first into $k_1$ groups that are each divided into $k_2$ groups and then into $k_3$ groups (with $k = k_1k_2k_3$ groups in total in the finest partition):
\begin{subequations}\label{eq:hier_SBM_affinity_general_3_level}
\begin{align*}
    \bm Q = \frac{1}{k} \left [\bm I_{k_1} \otimes \bm \omega_2  + d (\bm 1_{k_1}^{} \bm 1_{k_1}^\top -\bm I_{k_1}) \otimes \bm 1_{k_2k_3}^{} \bm 1_{k_2k_3}^\top \right ] \enspace ,\\      
    \text{where}\quad \bm \omega_2 = \bm I_{k_2} \otimes \bm \omega_1  + c (\bm 1_{k_2}^{} \bm 1_{k_2}^\top -\bm I_{k_2}) \otimes \bm 1_{k_3}^{} \bm 1_{k_3}^\top \enspace ,\\
    \text{and}\quad \bm \omega_1 = a \bm I_{k_3}  + b (\bm 1_{k_3}^{}\bm 1_{k_3}^\top - \bm I_{k_3}) \enspace ,
\end{align*}
\end{subequations}
where $a,b,c$ and $d$ are parameters that do not depend on $n$. 
The corresponding eigenvalues of $\bm Q$ now take four distinct values $\alpha_1, \alpha_2, \alpha_3$ and $\alpha_4$
\begin{align*}
\alpha_1 & = [a + (k_3-1)b + (k_2-1)k_3 c + (k_1-1)k_2k_3 d]/k   \\
  \alpha_2 & = [a + (k_3-1)b + (k_2-1)k_3 c - k_2k_3 d]/k  \\
  \alpha_3 & = [a + (k_3-1)b - k_3 c]/k \\
  \alpha_4 & = (a -b)/k \enspace , 
\end{align*}
which occur with the following multiplicities,
\begin{align*}
  \alpha_1: & \quad (\text{once}) \\
  \alpha_2: & \quad (k_1-1~\text{times}) \\
  \alpha_3: & \quad (k_1k_2-k_1)~\text{times}) \\
  \alpha_4: & \quad (k-k_1k_2~\text{times}) \enspace .
\end{align*}

Since all parameters are non-negative, $\alpha_1$ will again be have the largest magnitude. The order of the remaining eigenvalues will depend on the relative values of the density parameters $a,b,c$ and $d$:
\begin{align*}
  \alpha_2 - \alpha_3 & = \frac{c-d}{k_1} \\
  \alpha_3 - \alpha_4 & = \frac{b-c}{k_1k_2} \\
  \alpha_2 - \alpha_4 & = \frac{b+(k_2-1)c-k_2d}{k_1k_2} \enspace .
\end{align*}

\begin{figure*}
	\includegraphics[width=\linewidth]{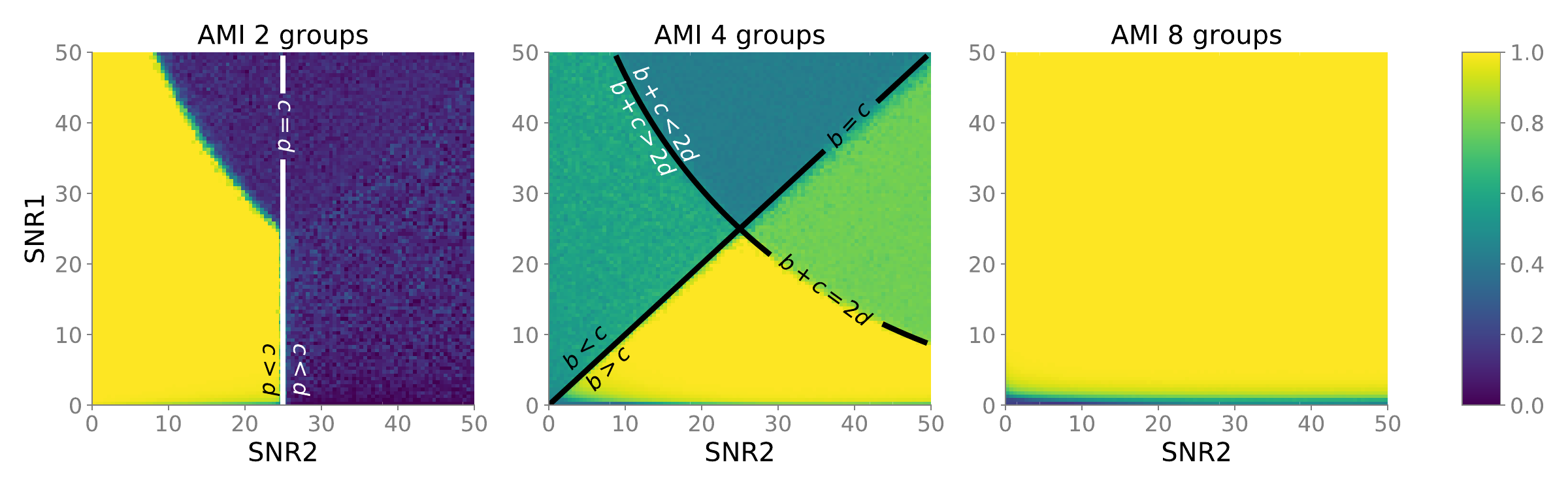}
	\includegraphics[width=\linewidth]{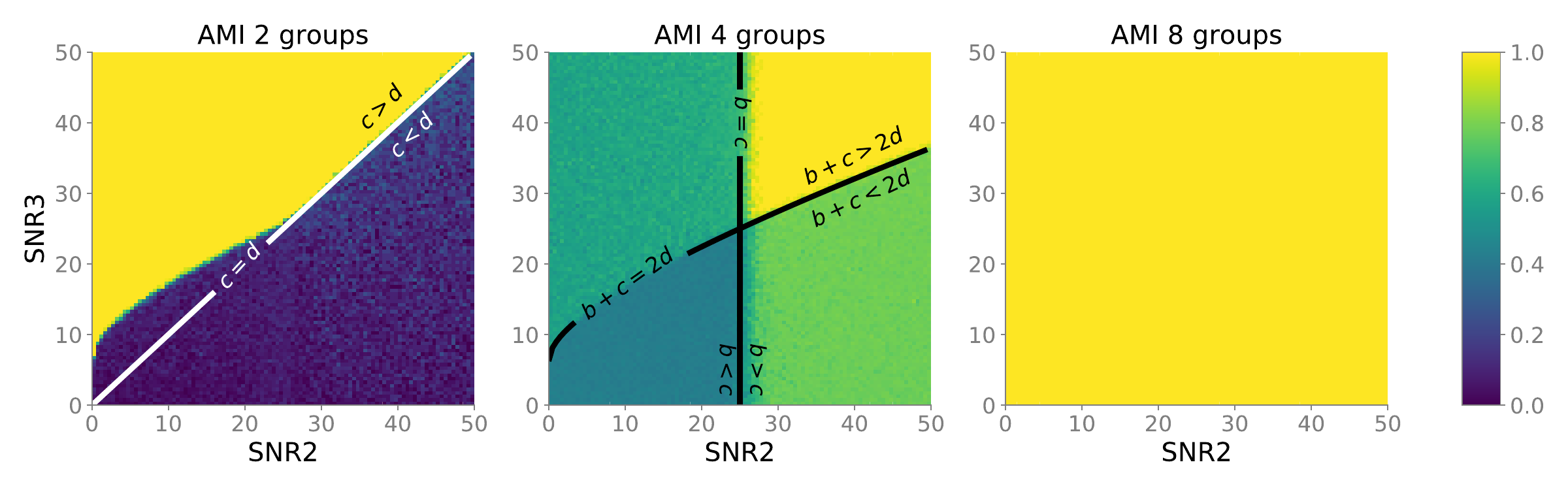}
	\caption{Detection of communities across three levels of hierarchy. At each level of the hierarchy the groups are bi-partitioned into two groups. \textit{Top:} We fix the SNR3 (coarsest partition, 2 groups) = 25 and vary SNR2 (4 groups) and SNR1 (finest partition, 8 groups) from 0 to 50. \textit{Bottom:} We fix the SNR1 (finest partition, 8 groups) = 25 and vary SNR2 (4 groups) and SNR3 (coarsest partition, 2 groups) from 0 to 50.}
	\label{fig:4}
\end{figure*}

In the prototypical hierarchical case in which communities become denser at finer resolutions, i.e., $a>b>c>d$, then $\alpha_1>\alpha_2>\alpha_3>\alpha_4$. 
Following our findings for two-level hierarchies we can expect that the reordering of the eigenvalues will lead to changes in the partitions detected at the coarse and intermediate levels. Therefore whenever $c<d$, $b<c$, and/or $b+(k_2-1)c<k_2d$ then there will exist an alternative partition that has a greater SNR.

We verify that these transitions occur in the three-level case with a similar experimental setup as we did for the two-level case, this time with $k_1=k_2=k_3=2$ and a four-parameter affinity matrix:
\begin{equation}\label{eq:hier_SBM_affinity_3level}
\bm \Omega = \frac{1}{n}
\begin{bmatrix} 
  a & b & c & c & d & d & d & d \\
  b & a & c & c & d & d & d & d \\
  c & c & a & b & d & d & d & d \\
  c & c & b & a & d & d & d & d \\
  d & d & d & d & a & b & c & c \\
  d & d & d & d & b & a & c & c \\
  d & d & d & d & c & c & a & b \\
  d & d & d & d & c & c & b & a 
\end{bmatrix} \enspace .
\end{equation}

We perform two sets of experiments in which we vary the SNR at each of the three levels. We refer to these as SNR1 (finest partition into 8 groups), SNR2 (intermediate partition into 4 groups) and SNR3 (coarsest partition into 2 groups). In the first experiment we fix $\text{SNR}3=25$ and generate hierarchical communities with SNR1 and SNR2 ranging from $0$ to $50$. Then in the second experiment we fix $\text{SNR}1=25$ and generate hierarchical communities with SNR2 and SNR3 ranging from $0$ to $50$. 

Figure~\ref{fig:4} shows the results of these experiments, the first experiment in the top row and the second experiment in the bottom row. We see that we can recover the finest partition in all cases as long as we are above the detectability limit. We can clearly see that the transitions in the recovery of the planted partitions closely matches the changes in the eigenvalues, indicated by the contour lines.

\section{Conclusion}
We have shown that the presence of multiple, consistent planted partitions can either enhance or diminish the detectability relative to networks with a single planted partition. 
Neither the general signal-to-noise ratio $\SNRl$, nor the individual ratios $\SNRc$ and $\SNRf$ are sufficient on their own to describe the detectability of hierarchical partitions.
Instead we must use these in combination with $\SNRcond$ in order to determine the detectability of a given level and consider the conditions under which the partition with the greatest SNR may change in non-identifiable cases~\cite{Schaub2020}. 
These non-identifiable cases further motivate methods for exploring presence of alternative ``good'' partitions in a network~\cite{kirkley2022representative,  mangold2023generative, Peele1602548, peixoto2021revealing}. 
While we consider the context of detecting hierarchical, our results should also be relevant for multi-resolution methods~\cite{jeub2018multiresolution}. 
In future we will investigate the effect of multiple inconsistent (i.e., non-nested) partitions.

\acknowledgments{MTS was supported by the Ministry of Culture and Science (MKW) of the German State of North Rhine-Westphalia (``NRW R\"uckkehrprogramm'').}

\bibliography{wns-bib}

\begin{thebibliography}{20}%
\makeatletter
\providecommand \@ifxundefined [1]{%
 \@ifx{#1\undefined}
}%
\providecommand \@ifnum [1]{%
 \ifnum #1\expandafter \@firstoftwo
 \else \expandafter \@secondoftwo
 \fi
}%
\providecommand \@ifx [1]{%
 \ifx #1\expandafter \@firstoftwo
 \else \expandafter \@secondoftwo
 \fi
}%
\providecommand \natexlab [1]{#1}%
\providecommand \enquote  [1]{``#1''}%
\providecommand \bibnamefont  [1]{#1}%
\providecommand \bibfnamefont [1]{#1}%
\providecommand \citenamefont [1]{#1}%
\providecommand \href@noop [0]{\@secondoftwo}%
\providecommand \href [0]{\begingroup \@sanitize@url \@href}%
\providecommand \@href[1]{\@@startlink{#1}\@@href}%
\providecommand \@@href[1]{\endgroup#1\@@endlink}%
\providecommand \@sanitize@url [0]{\catcode `\\12\catcode `\$12\catcode
  `\&12\catcode `\#12\catcode `\^12\catcode `\_12\catcode `\%12\relax}%
\providecommand \@@startlink[1]{}%
\providecommand \@@endlink[0]{}%
\providecommand \url  [0]{\begingroup\@sanitize@url \@url }%
\providecommand \@url [1]{\endgroup\@href {#1}{\urlprefix }}%
\providecommand \urlprefix  [0]{URL }%
\providecommand \Eprint [0]{\href }%
\providecommand \doibase [0]{http://dx.doi.org/}%
\providecommand \selectlanguage [0]{\@gobble}%
\providecommand \bibinfo  [0]{\@secondoftwo}%
\providecommand \bibfield  [0]{\@secondoftwo}%
\providecommand \translation [1]{[#1]}%
\providecommand \BibitemOpen [0]{}%
\providecommand \bibitemStop [0]{}%
\providecommand \bibitemNoStop [0]{.\EOS\space}%
\providecommand \EOS [0]{\spacefactor3000\relax}%
\providecommand \BibitemShut  [1]{\csname bibitem#1\endcsname}%
\let\auto@bib@innerbib\@empty
%</preamble>
\bibitem [{\citenamefont {Fortunato}(2010)}]{Fortunato2010}%
  \BibitemOpen
  \bibfield  {author} {\bibinfo {author} {\bibfnamefont {Santo}\ \bibnamefont
  {Fortunato}},\ }\bibfield  {title} {\enquote {\bibinfo {title} {{C}ommunity
  detection in graphs},}\ }\href {\doibase 10.1016/j.physrep.2009.11.002}
  {\bibfield  {journal} {\bibinfo  {journal} {Physics Reports}\ }\textbf
  {\bibinfo {volume} {486}},\ \bibinfo {pages} {75--174} (\bibinfo {year}
  {2010})}\BibitemShut {NoStop}%
\bibitem [{\citenamefont {Girvan}\ and\ \citenamefont
  {Newman}(2002)}]{girvan2002community}%
  \BibitemOpen
  \bibfield  {author} {\bibinfo {author} {\bibfnamefont {Michelle}\
  \bibnamefont {Girvan}}\ and\ \bibinfo {author} {\bibfnamefont {Mark~EJ}\
  \bibnamefont {Newman}},\ }\bibfield  {title} {\enquote {\bibinfo {title}
  {Community structure in social and biological networks},}\ }\href@noop {}
  {\bibfield  {journal} {\bibinfo  {journal} {Proceedings of the national
  academy of sciences}\ }\textbf {\bibinfo {volume} {99}},\ \bibinfo {pages}
  {7821--7826} (\bibinfo {year} {2002})}\BibitemShut {NoStop}%
\bibitem [{\citenamefont {Newman}\ and\ \citenamefont
  {Girvan}(2004)}]{Newman2004}%
  \BibitemOpen
  \bibfield  {author} {\bibinfo {author} {\bibfnamefont {M.~E.~J.}\
  \bibnamefont {Newman}}\ and\ \bibinfo {author} {\bibfnamefont
  {M.}~\bibnamefont {Girvan}},\ }\bibfield  {title} {\enquote {\bibinfo {title}
  {{F}inding and evaluating community structure in networks},}\ }\href
  {\doibase 10.1103/PhysRevE.69.026113} {\bibfield  {journal} {\bibinfo
  {journal} {Phys. Rev. E}\ }\textbf {\bibinfo {volume} {69}},\ \bibinfo
  {pages} {026113} (\bibinfo {year} {2004})}\BibitemShut {NoStop}%
\bibitem [{\citenamefont {Schaub}\ \emph {et~al.}(2017)\citenamefont {Schaub},
  \citenamefont {Delvenne}, \citenamefont {Rosvall},\ and\ \citenamefont
  {Lambiotte}}]{Schaub2017}%
  \BibitemOpen
  \bibfield  {author} {\bibinfo {author} {\bibfnamefont {Michael~T}\
  \bibnamefont {Schaub}}, \bibinfo {author} {\bibfnamefont {Jean-Charles}\
  \bibnamefont {Delvenne}}, \bibinfo {author} {\bibfnamefont {Martin}\
  \bibnamefont {Rosvall}}, \ and\ \bibinfo {author} {\bibfnamefont {Renaud}\
  \bibnamefont {Lambiotte}},\ }\bibfield  {title} {\enquote {\bibinfo {title}
  {The many facets of community detection in complex networks},}\ }\href@noop
  {} {\bibfield  {journal} {\bibinfo  {journal} {Applied Network Science}\
  }\textbf {\bibinfo {volume} {2}},\ \bibinfo {pages} {4} (\bibinfo {year}
  {2017})}\BibitemShut {NoStop}%
\bibitem [{\citenamefont {Holland}\ \emph {et~al.}(1983)\citenamefont
  {Holland}, \citenamefont {Laskey},\ and\ \citenamefont
  {Leinhardt}}]{holland1983stochastic}%
  \BibitemOpen
  \bibfield  {author} {\bibinfo {author} {\bibfnamefont {Paul~W}\ \bibnamefont
  {Holland}}, \bibinfo {author} {\bibfnamefont {Kathryn~Blackmond}\
  \bibnamefont {Laskey}}, \ and\ \bibinfo {author} {\bibfnamefont {Samuel}\
  \bibnamefont {Leinhardt}},\ }\bibfield  {title} {\enquote {\bibinfo {title}
  {Stochastic blockmodels: First steps},}\ }\href@noop {} {\bibfield  {journal}
  {\bibinfo  {journal} {Soc. Networks}\ }\textbf {\bibinfo {volume} {5}},\
  \bibinfo {pages} {109--137} (\bibinfo {year} {1983})}\BibitemShut {NoStop}%
\bibitem [{\citenamefont {Abbe}(2018)}]{abbe2018community}%
  \BibitemOpen
  \bibfield  {author} {\bibinfo {author} {\bibfnamefont {Emmanuel}\
  \bibnamefont {Abbe}},\ }\bibfield  {title} {\enquote {\bibinfo {title}
  {Community detection and stochastic block models: Recent developments},}\
  }\href@noop {} {\bibfield  {journal} {\bibinfo  {journal} {Journal of Machine
  Learning Research}\ }\textbf {\bibinfo {volume} {18}},\ \bibinfo {pages}
  {1--86} (\bibinfo {year} {2018})}\BibitemShut {NoStop}%
\bibitem [{\citenamefont {Decelle}\ \emph {et~al.}(2011)\citenamefont
  {Decelle}, \citenamefont {Krzakala}, \citenamefont {Moore},\ and\
  \citenamefont {Zdeborov{\'a}}}]{decelle2011asymptotic}%
  \BibitemOpen
  \bibfield  {author} {\bibinfo {author} {\bibfnamefont {Aurelien}\
  \bibnamefont {Decelle}}, \bibinfo {author} {\bibfnamefont {Florent}\
  \bibnamefont {Krzakala}}, \bibinfo {author} {\bibfnamefont {Cristopher}\
  \bibnamefont {Moore}}, \ and\ \bibinfo {author} {\bibfnamefont {Lenka}\
  \bibnamefont {Zdeborov{\'a}}},\ }\bibfield  {title} {\enquote {\bibinfo
  {title} {Asymptotic analysis of the stochastic block model for modular
  networks and its algorithmic applications},}\ }\href@noop {} {\bibfield
  {journal} {\bibinfo  {journal} {Physical Review E}\ }\textbf {\bibinfo
  {volume} {84}},\ \bibinfo {pages} {066106} (\bibinfo {year}
  {2011})}\BibitemShut {NoStop}%
\bibitem [{\citenamefont {Massouli{\'e}}(2014)}]{massoulie2014community}%
  \BibitemOpen
  \bibfield  {author} {\bibinfo {author} {\bibfnamefont {Laurent}\ \bibnamefont
  {Massouli{\'e}}},\ }\bibfield  {title} {\enquote {\bibinfo {title} {Community
  detection thresholds and the weak ramanujan property},}\ }in\ \href@noop {}
  {\emph {\bibinfo {booktitle} {Proceedings of the forty-sixth annual ACM
  symposium on Theory of computing}}}\ (\bibinfo {year} {2014})\ pp.\ \bibinfo
  {pages} {694--703}\BibitemShut {NoStop}%
\bibitem [{\citenamefont {Mossel}\ \emph {et~al.}(2018)\citenamefont {Mossel},
  \citenamefont {Neeman},\ and\ \citenamefont {Sly}}]{Mossel2018}%
  \BibitemOpen
  \bibfield  {author} {\bibinfo {author} {\bibfnamefont {Elchanan}\
  \bibnamefont {Mossel}}, \bibinfo {author} {\bibfnamefont {Joe}\ \bibnamefont
  {Neeman}}, \ and\ \bibinfo {author} {\bibfnamefont {Allan}\ \bibnamefont
  {Sly}},\ }\bibfield  {title} {\enquote {\bibinfo {title} {A proof of the
  block model threshold conjecture},}\ }\href {\doibase
  10.1007/s00493-016-3238-8} {\bibfield  {journal} {\bibinfo  {journal}
  {Combinatorica}\ }\textbf {\bibinfo {volume} {38}},\ \bibinfo {pages}
  {665--708} (\bibinfo {year} {2018})}\BibitemShut {NoStop}%
\bibitem [{\citenamefont {Mossel}\ \emph {et~al.}(2015)\citenamefont {Mossel},
  \citenamefont {Neeman},\ and\ \citenamefont
  {Sly}}]{mossel2015reconstruction}%
  \BibitemOpen
  \bibfield  {author} {\bibinfo {author} {\bibfnamefont {Elchanan}\
  \bibnamefont {Mossel}}, \bibinfo {author} {\bibfnamefont {Joe}\ \bibnamefont
  {Neeman}}, \ and\ \bibinfo {author} {\bibfnamefont {Allan}\ \bibnamefont
  {Sly}},\ }\bibfield  {title} {\enquote {\bibinfo {title} {Reconstruction and
  estimation in the planted partition model},}\ }\href@noop {} {\bibfield
  {journal} {\bibinfo  {journal} {Probability Theory and Related Fields}\
  }\textbf {\bibinfo {volume} {162}},\ \bibinfo {pages} {431--461} (\bibinfo
  {year} {2015})}\BibitemShut {NoStop}%
\bibitem [{\citenamefont {Moore}(2017)}]{moore2017computer}%
  \BibitemOpen
  \bibfield  {author} {\bibinfo {author} {\bibfnamefont {Cristopher}\
  \bibnamefont {Moore}},\ }\bibfield  {title} {\enquote {\bibinfo {title} {The
  computer science and physics of community detection: Landscapes, phase
  transitions, and hardness},}\ }\href@noop {} {\bibfield  {journal} {\bibinfo
  {journal} {arXiv preprint arXiv:1702.00467}\ } (\bibinfo {year}
  {2017})}\BibitemShut {NoStop}%
\bibitem [{\citenamefont {Abbe}\ and\ \citenamefont
  {Sandon}(2015)}]{abbe2015detection}%
  \BibitemOpen
  \bibfield  {author} {\bibinfo {author} {\bibfnamefont {Emmanuel}\
  \bibnamefont {Abbe}}\ and\ \bibinfo {author} {\bibfnamefont {Colin}\
  \bibnamefont {Sandon}},\ }\bibfield  {title} {\enquote {\bibinfo {title}
  {Detection in the stochastic block model with multiple clusters: proof of the
  achievability conjectures, acyclic bp, and the information-computation
  gap},}\ }\href@noop {} {\bibfield  {journal} {\bibinfo  {journal} {arXiv
  preprint arXiv:1512.09080}\ } (\bibinfo {year} {2015})}\BibitemShut {NoStop}%
\bibitem [{\citenamefont {Saade}\ \emph {et~al.}(2014)\citenamefont {Saade},
  \citenamefont {Krzakala},\ and\ \citenamefont {Zdeborov\'{a}}}]{Saade2014}%
  \BibitemOpen
  \bibfield  {author} {\bibinfo {author} {\bibfnamefont {Alaa}\ \bibnamefont
  {Saade}}, \bibinfo {author} {\bibfnamefont {Florent}\ \bibnamefont
  {Krzakala}}, \ and\ \bibinfo {author} {\bibfnamefont {Lenka}\ \bibnamefont
  {Zdeborov\'{a}}},\ }\bibfield  {title} {\enquote {\bibinfo {title} {Spectral
  clustering of graphs with the {B}ethe hessian},}\ }in\ \href
  {http://papers.nips.cc/paper/5520-spectral-clustering-of-graphs-with-the-bethe-hessian.pdf}
  {\emph {\bibinfo {booktitle} {Advances in Neural Information Processing
  Systems 27}}}\ (\bibinfo {year} {2014})\ pp.\ \bibinfo {pages}
  {406--414}\BibitemShut {NoStop}%
\bibitem [{\citenamefont {Vinh}\ \emph {et~al.}(2010)\citenamefont {Vinh},
  \citenamefont {Epps},\ and\ \citenamefont {Bailey}}]{Vinh2010}%
  \BibitemOpen
  \bibfield  {author} {\bibinfo {author} {\bibfnamefont {Nguyen~Xuan}\
  \bibnamefont {Vinh}}, \bibinfo {author} {\bibfnamefont {Julien}\ \bibnamefont
  {Epps}}, \ and\ \bibinfo {author} {\bibfnamefont {James}\ \bibnamefont
  {Bailey}},\ }\bibfield  {title} {\enquote {\bibinfo {title} {Information
  theoretic measures for clusterings comparison: Variants, properties,
  normalization and correction for chance},}\ }\href@noop {} {\bibfield
  {journal} {\bibinfo  {journal} {Journal of Machine Learning Research}\
  }\textbf {\bibinfo {volume} {11}},\ \bibinfo {pages} {2837--2854} (\bibinfo
  {year} {2010})}\BibitemShut {NoStop}%
\bibitem [{\citenamefont {Schaub}\ \emph {et~al.}(2023)\citenamefont {Schaub},
  \citenamefont {Li},\ and\ \citenamefont {Peel}}]{Schaub2020}%
  \BibitemOpen
  \bibfield  {author} {\bibinfo {author} {\bibfnamefont {Michael~T}\
  \bibnamefont {Schaub}}, \bibinfo {author} {\bibfnamefont {Jiaze}\
  \bibnamefont {Li}}, \ and\ \bibinfo {author} {\bibfnamefont {Leto}\
  \bibnamefont {Peel}},\ }\bibfield  {title} {\enquote {\bibinfo {title}
  {Hierarchical community structure in networks},}\ }\href@noop {} {\bibfield
  {journal} {\bibinfo  {journal} {Physical Review E}\ }\textbf {\bibinfo
  {volume} {107}},\ \bibinfo {pages} {054305} (\bibinfo {year}
  {2023})}\BibitemShut {NoStop}%
\bibitem [{\citenamefont {Kirkley}\ and\ \citenamefont
  {Newman}(2022)}]{kirkley2022representative}%
  \BibitemOpen
  \bibfield  {author} {\bibinfo {author} {\bibfnamefont {Alec}\ \bibnamefont
  {Kirkley}}\ and\ \bibinfo {author} {\bibfnamefont {Mark~E.J.}\ \bibnamefont
  {Newman}},\ }\bibfield  {title} {\enquote {\bibinfo {title} {Representative
  community divisions of networks},}\ }\href@noop {} {\bibfield  {journal}
  {\bibinfo  {journal} {Communications Physics}\ }\textbf {\bibinfo {volume}
  {5}},\ \bibinfo {pages} {40} (\bibinfo {year} {2022})}\BibitemShut {NoStop}%
\bibitem [{\citenamefont {Mangold}\ and\ \citenamefont
  {Roth}(2023)}]{mangold2023generative}%
  \BibitemOpen
  \bibfield  {author} {\bibinfo {author} {\bibfnamefont {Lena}\ \bibnamefont
  {Mangold}}\ and\ \bibinfo {author} {\bibfnamefont {Camille}\ \bibnamefont
  {Roth}},\ }\bibfield  {title} {\enquote {\bibinfo {title} {Generative models
  for two-ground-truth partitions in networks},}\ }\href@noop {} {\bibfield
  {journal} {\bibinfo  {journal} {Physical Review E}\ }\textbf {\bibinfo
  {volume} {108}},\ \bibinfo {pages} {054308} (\bibinfo {year}
  {2023})}\BibitemShut {NoStop}%
\bibitem [{\citenamefont {Peel}\ \emph {et~al.}(2017)\citenamefont {Peel},
  \citenamefont {Larremore},\ and\ \citenamefont {Clauset}}]{Peele1602548}%
  \BibitemOpen
  \bibfield  {author} {\bibinfo {author} {\bibfnamefont {Leto}\ \bibnamefont
  {Peel}}, \bibinfo {author} {\bibfnamefont {Daniel~B.}\ \bibnamefont
  {Larremore}}, \ and\ \bibinfo {author} {\bibfnamefont {Aaron}\ \bibnamefont
  {Clauset}},\ }\bibfield  {title} {\enquote {\bibinfo {title} {The ground
  truth about metadata and community detection in networks},}\ }\href {\doibase
  10.1126/sciadv.1602548} {\bibfield  {journal} {\bibinfo  {journal} {Science
  Advances}\ }\textbf {\bibinfo {volume} {3}} (\bibinfo {year} {2017}),\
  10.1126/sciadv.1602548}\BibitemShut {NoStop}%
\bibitem [{\citenamefont {Peixoto}(2021)}]{peixoto2021revealing}%
  \BibitemOpen
  \bibfield  {author} {\bibinfo {author} {\bibfnamefont {Tiago~P}\ \bibnamefont
  {Peixoto}},\ }\bibfield  {title} {\enquote {\bibinfo {title} {Revealing
  consensus and dissensus between network partitions},}\ }\href@noop {}
  {\bibfield  {journal} {\bibinfo  {journal} {Physical Review X}\ }\textbf
  {\bibinfo {volume} {11}},\ \bibinfo {pages} {021003} (\bibinfo {year}
  {2021})}\BibitemShut {NoStop}%
\bibitem [{\citenamefont {Jeub}\ \emph {et~al.}(2018)\citenamefont {Jeub},
  \citenamefont {Sporns},\ and\ \citenamefont
  {Fortunato}}]{jeub2018multiresolution}%
  \BibitemOpen
  \bibfield  {author} {\bibinfo {author} {\bibfnamefont {Lucas~GS}\
  \bibnamefont {Jeub}}, \bibinfo {author} {\bibfnamefont {Olaf}\ \bibnamefont
  {Sporns}}, \ and\ \bibinfo {author} {\bibfnamefont {Santo}\ \bibnamefont
  {Fortunato}},\ }\bibfield  {title} {\enquote {\bibinfo {title}
  {Multiresolution consensus clustering in networks},}\ }\href@noop {}
  {\bibfield  {journal} {\bibinfo  {journal} {Scientific reports}\ }\textbf
  {\bibinfo {volume} {8}},\ \bibinfo {pages} {3259} (\bibinfo {year}
  {2018})}\BibitemShut {NoStop}%
\end{thebibliography}%

\end{document}